# SOUL: An Energy-Efficient Unsupervised Online Learning Seizure Detection Classifier

Adelson Chua, *Graduate Student Member, IEEE*, Michael I. Jordan, *Fellow, IEEE*, and Rikky Muller, *Senior Member, IEEE*

*Abstract*—Implantable devices that record neural activity and detect seizures have been adopted to issue warnings or trigger neurostimulation to suppress epileptic seizures. Typical seizure detection systems rely on high-accuracy offline-trained machine learning classifiers that require manual retraining when seizure patterns change over long periods of time. For an implantable seizure detection system, a low-power, at-the-edge, online learning algorithm can be employed to dynamically adapt to the neural signal drifts, thereby maintaining high accuracy without external intervention. This work proposes SOUL: Stochastic-gradient-descent-based Online Unsupervised Logistic regression classifier. After an initial offline training phase, continuous online unsupervised classifier updates are applied *in situ*, which improves sensitivity in patients with drifting seizure features. SOUL was tested on two human electroencephalography (EEG) datasets: the Children's Hospital Boston and the Massachusetts Institute of Technology (CHB-MIT) scalp EEG dataset and a long (>100 h) intracranial EEG dataset. It was able to achieve an average sensitivity of 97.5% and 97.9% for the two datasets, respectively, at >95% specificity. Sensitivity improved by at most 8.2% on long-term data when compared to a typical seizure detection classifier. SOUL was fabricated in Taiwan Semiconductor Manufacturing Company (TSMC's) 28 nm process occupying 0.1 mm$^2$ and achieves 1.5 nJ/classification energy efficiency, which is at least 24× more efficient than state-of-the-art.

## I. INTRODUCTION

EPILEPSY is a serious neurological disorder affecting around 50 million people worldwide [1] and is usually characterized by recurrent seizures. Seizure frequency varies greatly from person to person and can severely impact a person's quality of life. Seizures are often seen as a series of high-amplitude, high-frequency electrical signals [2], [3] that can be measured through electroencephalography (EEG).

As an aid to patients suffering from seizures, advisory systems have been developed that warn patients when a seizure is about to occur. The system first records EEG signals, either intracranially or on the scalp, and then performs classification to detect the onset or presence of a seizure. Closed-loop implantable neuromodulators have also been deployed for seizure treatment. These systems detect seizure events within an acceptable latency (typically <5 s [3], [4]) and trigger neurostimulation to suppress the seizure. The NeuroPace Responsive neurostimulation (RNS) [2]–[5] and the Medtronic Deep-brain stimulation (DBS) [6], [7] are two medically approved devices of this kind. These devices utilize a small, battery-powered pulse generator surgically implanted in the skull with two electrode leads that are implanted intracranially and/or epicortically. This treatment method has demonstrated clinical efficacy in terms of reducing long-term seizure occurrence, reporting a reduction of 66% of seizures by Year 6 for the NeuroPace RNS [2] and 75% median reduction of seizures by Year 7 for the Medtronic DBS [7].

In some patients, EEG seizure patterns can change over time, which can be due to electrode displacement and impedance changes [8] or shifts to the patients' circadian profiles [9]. Such changes would require regular retraining of the seizure detection algorithms to maintain high accuracy detection. Prior art in seizure prediction and detection utilized long-term datasets to capture such variations [10], resulting in seizure detection accuracy greater than 90%. However, the classifier algorithms in [10] were software-only implementations, where computational complexity and memory requirements were not a design consideration. For implantable closed-loop seizure detection systems, energy efficiency, area utilization, and long-term accuracy become important design constraints.

There were several application-specific integrated circuits (ASICs) for the state-of-the-art seizure event classification published in the literature. These on-chip classifiers typically employ support vector machines (SVMs) [11]–[16] due to their high accuracy and relatively simple implementations. However, these classifiers can still have significant memory requirements to hold their support vectors. At least 64 kB [11]–[16] of memory is required leading to high on-chip area and power consumption. Moreover, while these ASIC seizure detectors usually incorporate on-chip feature calculations, training and

Manuscript received October 1, 2021; revised January 5, 2022 and March 14, 2022; accepted April 21, 2022. This article was approved by Associate Editor Nick van Helleputte. This work was supported in part by the National Science Foundation (NSF) Faculty Early Career Development Program (CAREER) under Grant 1847710, in part by the Wagner Foundation, in part by the Weill Neurohub, in part the Army Research Office under Contract W911NF-16-1-0368, in part by the Department of Science and Technology - Science Education Institute (DOST-SEI), and in part by the University of the Philippines Diliman. *(Corresponding author: Adelson Chua.)*

Adelson Chua is with the Department of Electrical Engineering and Computer Sciences, University of California at Berkeley, Berkeley, CA 94720 USA, and also with the Electrical and Electronics Engineering Institute, University of the Philippines Diliman, Quezon City 1101, Philippines (e-mail: adelson.chua@berkeley.edu).

Michael I. Jordan is with the Department of Electrical Engineering and Computer Sciences and the Department of Statistics, University of California at Berkeley, Berkeley, CA 94720 USA (e-mail: jordan@cs.berkeley.edu).

Rikky Muller is with the Department of Electrical Engineering and Computer Sciences, University of California at Berkeley, Berkeley, CA 94720 USA, and also with Chan-Zuckerberg Biohub, San Francisco, CA 94158 USA (e-mail: rikky@berkeley.edu).

Color versions of one or more figures in this article are available at https://doi.org/10.1109/JSSC.2022.3172231.

Digital Object Identifier 10.1109/JSSC.2022.3172231

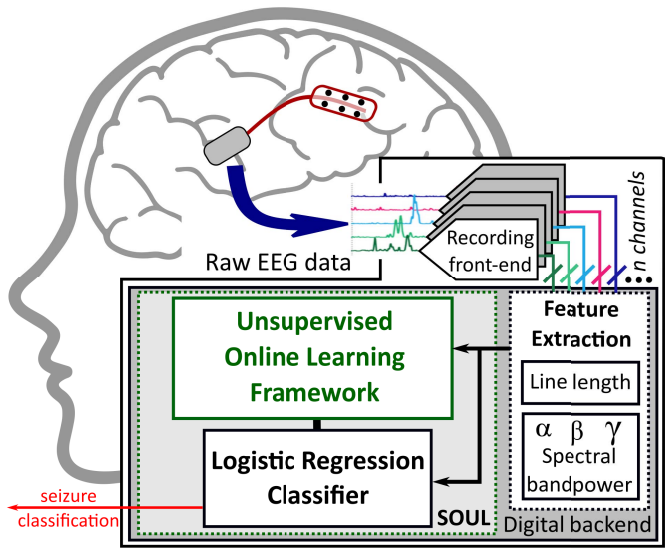

Fig. 1. Proposed seizure detection system featuring a fully unsupervised online learning framework to maintain long-term high accuracy detection.

its associated computational complexity is usually completely offloaded to software. The calculated SVM model parameters after offline training are loaded into the chip, which then performs online seizure classification. Consequently, for a seizure detection system to remain accurate over long periods of time on patients with changing seizure patterns, regular signal post-processing, labeling, and retraining by an expert physician would be required. Such external intervention can also be costly and impractical.

This work demonstrates the use of unsupervised online learning to dynamically adapt to changes in neural signal patterns over time and maintain high detection accuracy without external intervention. This will be referred to as Stochastic-gradient-descent-based Online Unsupervised Logistic regression classifier (SOUL) [17], [18], shown in Fig. 1. SOUL is initially trained offline and then feature weights are updated *in situ*. Moreover, due to the computationally simple algorithm and architectural optimizations used, SOUL is significantly more energy efficient than the state-of-the-art on-chip seizure detectors.

The remainder of this article is organized as follows. Section II describes the proposed unsupervised online learning classifier for seizure detection. Section III evaluates the performance of the proposed classifier on two human EEG datasets: the Children's Hospital Boston and the Massachusetts Institute of Technology (CHB-MIT) scalp EEG dataset [19] and the intracranial EEG (iEEG) dataset collected from a study conducted by the University of Melbourne [8]. Section IV describes the hardware architecture for the proposed algorithm and the energy efficiency measurements of the fabricated chip. Section V presents a discussion of the results and provides comparisons with prior work. Section VI concludes the article.

## II. ALGORITHM DESIGN

### A. State-of-the-Art Online Learning Seizure Detectors

Gradient descent is a commonly used optimization algorithm for training machine learning model parameters. The algorithm involves several matrix-based operations, which require mathematically complex calculations that lead to high-energy per classification when implemented in hardware. For SVM-based classifiers, only [15] has demonstrated online learning through gradient descent on an ASIC. When the system was run on a 6-h EEG recording, the sensitivity was maintained at 96.1% while the false alarm rate was reduced from 1.83% to 0.34%. However, the reported power consumption was on the order of milliwatts, with energy efficiency at 170 $\mu$J/classification.

A simpler approximation of SVM online learning, not using gradient descent, has also been demonstrated [16]. The implementation incorporates a post-processing block that selectively replaces the preloaded support vectors on-chip. It has reported a sensitivity improvement from 39.5% to 71.9% on a single patient. The energy efficiency of the classifier was reported to be at 0.34 $\mu$J/classification, which is 500× more efficient than the previous work utilizing gradient descent. An online learning neural network-based classifier has also been demonstrated [20] utilizing a post-processing block to modify the detection threshold for the classifier. A 4.2% improvement in accuracy attributed to the online learning scheme was reported on a 40-min-long EEG dataset. Energy efficiency was measured to be 2.06 $\mu$J/classification, which is 6× greater than [16] due to the complexity of the classifier.

A common limitation of these implementations is that the online learning processes are entirely supervised. That is, these require external labels for the algorithms to work. Moreover, these were not tested on long-term continuous datasets (>100 h/patient). Algorithms that work well on short EEG recordings can fail to work on longer recordings. Gradient-descent-based online learning can work in the long term since it is an iterative optimization algorithm. However, its computational complexity leads to high-power consumption for complex classifiers such as SVMs.

This work proposes the use of a binary classifier based on a generalized linear model, such as logistic regression, which can leverage gradient descent as an optimization algorithm without significant complexity. Moreover, by utilizing the classifier's output as the training label, unsupervised online learning can be achieved.

### B. Logistic Regression as a Classifier

Logistic regression is a probabilistic model that utilizes the logistic function to map the weighted linear combination of input features to real values between 0 and 1, which can then be interpreted as probabilities. Thresholding the output to any value between 0 and 1 (typically 0.5), would result in binary classification. The standard logistic function is shown in the following equation:

$$p(w_t, x_t) = \frac{1}{1 + e^{-w_t^T x_t}}. \quad (1)$$

The $w_t$ term refers to the vector of logistic regression feature weights corresponding to the vector of feature inputs $x_t$. The values for the weights in $w_t$ are calculated through an iterative process to best fit the logistic function on the labeled set of feature inputs. That iterative process is how the logistic regression classifier is trained and is typically done in software as it uses the gradient descent algorithm for the optimization.

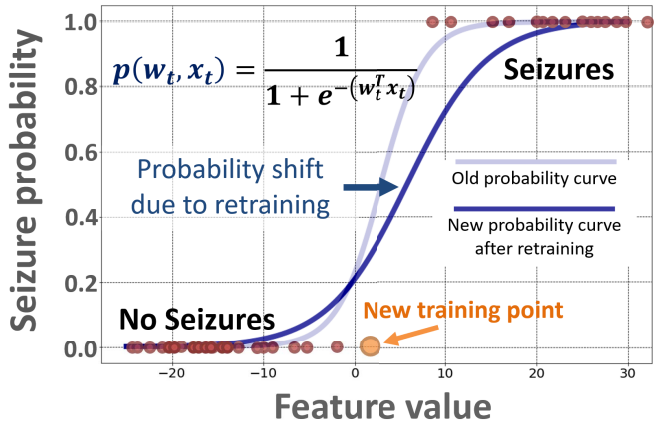

Fig. 2. Logistic function shift toward a new optimal curve due to updated feature weights after logistic regression retraining. The function shifts to the right (darker line) due to the introduction of a new training point (enlarged circle).

Since the output of logistic regression depends on the linear combination of weights and features, it performs very well on linearly separable data. As seizure and nonseizure events usually exhibit this property (especially using features that can detect the high amplitude and high-frequency seizure signals, more on this later this section), logistic regression can be used as a seizure event classifier. Prior work has compared logistic regression against other classifiers [23]–[25] for this application and has shown comparable performance. However, when the feature values between seizures and nonseizures vary over time, linear separability between the two classes cannot be maintained, leading to accuracy degradation. SVMs, on the other hand, can utilize nonlinear kernel functions to force class separability leading to better accuracy. This is the reason why the state-of-the-art seizure classifiers typically use SVMs.

The limitation of logistic regression on the diminishing linear separability can be mitigated if logistic regression can track feature value changes over time. As the feature values $x_t$ drift, the optimal feature weights $w_t$ that were calculated during offline training might not hold true anymore. Thus, if a new set of weights can be calculated beyond the initial offline training period (i.e., online), the logistic function can shift dynamically, to maintain optimality. Fig. 2 illustrates this function shift on a 1-D feature example. As a new training point is introduced, the curve shifts to the right to ensure that the new point is properly classified.

As will be discussed in the succeeding sections, SOUL calculates four different features per channel over eight total channels, resulting in 32 total features per sample. In this case, the logistic function is plotted on a 32-D space. The logistic function shift corresponds to a calculation of a new vector of 32 feature weights $w_t$ that will change the classifier's decision boundary in the 32-D space.

### C. Stochastic Gradient Descent for Logistic Regression

The stochastic gradient descent (SGD) algorithm is an iterative method of optimizing the classifier feature weights by approximating the calculation of the gradient descent using a new set of feature inputs [26]. This algorithm avoids the complex computation of the gradient on the whole training data. While SGD is an approximation, it can be used to dynamically update the feature weights online through a defined optimization algorithm. Fig. 3 describes the procedure. A set of feature weights are initially trained offline, using any software-based training algorithm available, to achieve the best possible accuracy from the training data. This process provides a good baseline for logistic regression classification. Cross-validation is performed by running the classification and SGD on validation data without external labels. This is where different hyperparameters are tuned to maximize accuracy during the unsupervised online learning phase. Then, upon classifier deployment, the classifier can utilize the test data to update the feature weights using SGD on chip.

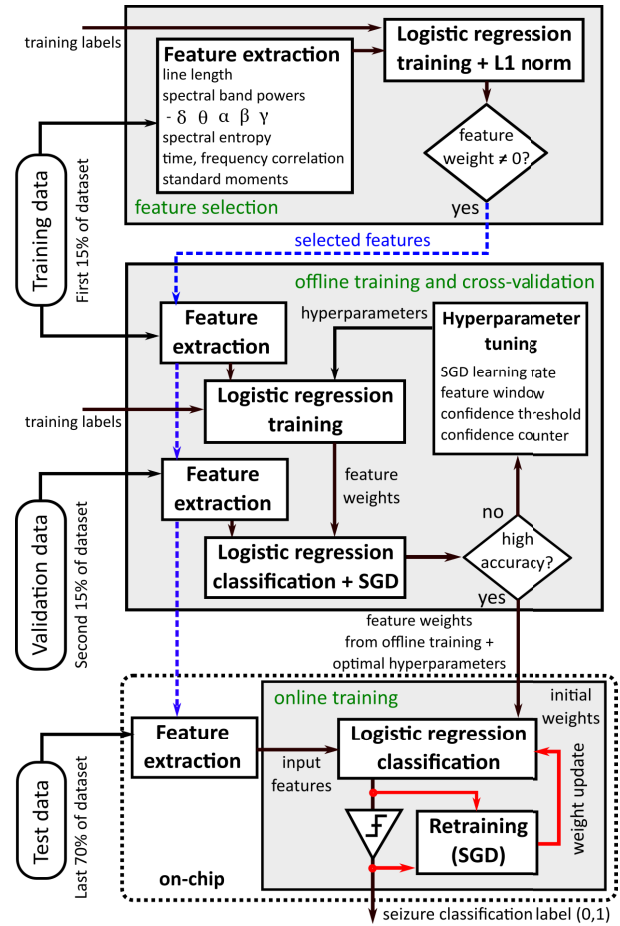

Fig. 3. Feature selection, offline training and cross-validation, and online (on-chip) retraining scheme. Feature selection reduces the features to be extracted through $L_1$-norm regularization. Offline training phase generates the best possible set of starting feature weights and hyperparameter values for on-chip classification. Feature weights are dynamically updated on chip using SGD (highlighted in red).

The logistic regression weight update is computationally simple as shown in the following equation:

$$w_{t+1} = w_t + \eta(y_t - p(w_t, x_t))x_t. \quad (2)$$

The $w_{t+1}$ term refers to the next set of feature weights after the update; $\eta$ is the learning rate of the algorithm, which controls how much the feature weights will change based on new data; and $y_t$ is the corresponding label for the current feature input. The SGD-based feature weight update can be done in a single iteration with minimal hardware. The update is also done in one epoch (i.e., one-shot retraining on the new data) as SOUL does not save the previous training points to minimize the memory requirements. The logistic function calculation can also be implemented using a lookup table (LUT) to further reduce the computational complexity. Architectural optimizations will be covered further in Section IV.

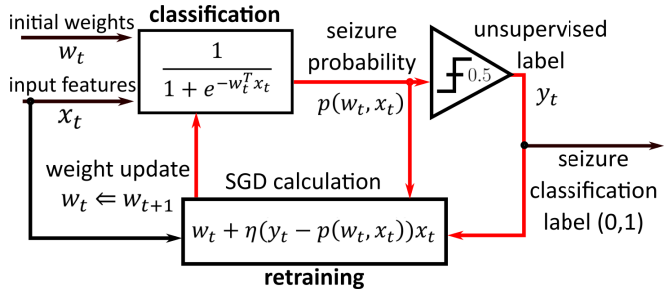

Fig. 4. Feedback loop when using the classifier's own output probability (rounded off to 0 or 1) as the training label for SGD.

### D. Enabling Unsupervised Learning

Traditionally, SGD is meant for supervised learning, where an external label is provided for every data input [26]. However, for an implantable system operating *in situ*, externally provided labels are not readily available. Thus, this approach places SGD within an unsupervised learning paradigm during the online classification phase. This is implemented through bootstrapping, which uses the classifier's predicted probability output to update its own feature weights. The classifier's output probability $p(w_t, x_t)$ is rounded to either 0 or 1 and is then treated as a label $y_t$ for SGD. This creates a positive feedback path between the classifier's output and its input training label, highlighted in Fig. 4. Consequently, the cumulative accuracy over time is heavily dependent on the initial classifier accuracy after the offline training phase. It is critical that the initial logistic regression weights achieve a high classification accuracy during the training phase. The feature set used for this work, which will be described later in this section, adequately separates seizure and nonseizure events. Therefore, achieving high classification accuracy, at least during the training phase, is possible. The unsupervised online learning classifier (SOUL) will be tracking the long-term changes in these seizure and nonseizure patterns through feature weight updates *in situ*.

### E. Making the Unsupervised Online Learning Robust

While high classification accuracy is required for offline training, the classifier can still make occasional errors. Generally, any misclassification can degrade the accuracy due to the positive feedback, as the classifier will retrain in the wrong direction. To avoid such an occurrence, the weights are only updated once a specified confidence threshold (CT) is reached by the logistic function output. Moreover, a series of high-confidence (HC) predictions are required to trigger the online feature weight update, shown in Fig. 5. Only the last set of features after the consecutive HC predictions would then be used as new data for the SGD algorithm on the next weight update. This process ensures that short-term misclassifications and glitches will not negatively affect the update. The CT and HC counter become additional hyperparameters during the offline training phase (Fig. 3) and are tuned on a patient-specific basis.

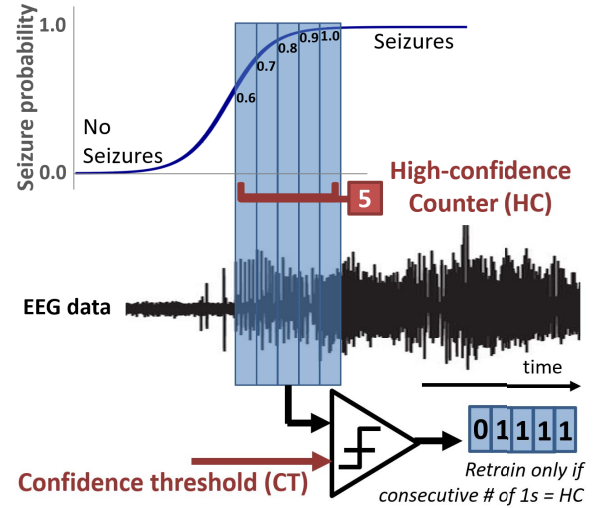

Fig. 5. Confidence thresholding technique implemented to only train the classifier once a series of HC predictions are generated from the logistic regression output.

### F. Classifier Features

The feature extraction unit computes two main feature classes (Fig. 1): line length and spectral band powers for three frequency bands. These features are commonly used in seizure detection systems since they capture amplitude and frequency-dependent patterns usually attributed to seizure events. Other features were also considered, such as spectral entropy and time/frequency correlations, but were down-selected after running the initial training with $L_1$-norm regularization (Fig. 3), which zeroed out most of these features. The same $L_1$-norm penalization was done to remove highly correlated channels which eventually lead to the current eight channels supported by the classifier. After the channel and feature selection processes, the classification accuracies only decreased by <2% on average relative to the accuracies on the original set of features on all channels included in the datasets.

Line length [27] captures the high amplitude and high-frequency data characteristic of seizures, defined by the sum of the absolute value of differences between consecutive points, as shown in the following equation:

$$\sum_{t=1}^{N-1} |x_t - x_{t-1}|. \tag{3}$$

Spectral band power captures frequency-dependent patterns, calculated by summing the spectral power over a specific frequency band. This feature has been shown to separate seizure and nonseizure events very well [13]. This can also be approximated by passing the signal through a bandpass filter on a specified frequency range and then performing a sum of squares, exploiting Parseval's theorem, as shown in (4). This approximation eliminates the need for dedicated fast Fourier transform (FFT) hardware in the system

$$\sum_{t=0}^{N-1} x_t^2 = \frac{1}{N}\sum_{f=0}^{N-1}|X_f|^2. \quad (4)$$

The spectral band power is calculated for three EEG frequency bands: $\alpha$, 8–16 Hz; $\beta$, 16–32 Hz; $\gamma$, 32–96 Hz. Spectral power for the lower frequency bands were removed after the feature selection process described previously.

Both line length and spectral band power features require a specific sample window $N$. For this work, a 0.1 s window was used, which translates to a 100-sample sliding window for a 1 kHz input sampling rate. This feature window size was determined from the offline training phase (Fig. 3), as part of the hyperparameters that were optimized. This window controls how much input signal noise is smoothed out during the feature extraction process, which tends to dampen the feature value response due to averaging. However, the response to sudden signal transitions (which can be indicative of seizures) can also be delayed. To capture such changes, a 99% feature window overlap was chosen. That is, a new feature is calculated for every sample, for an effective classification rate of 1 kHz.

## III. Classifier Performance

The performance of SOUL is tested using the iEEG dataset [8] and the CHB-MIT scalp EEG dataset [19]. The former features >100 h recordings on three patients to demonstrate how online learning performs over a long period of time. The latter is a collection of relatively short recordings on 24 patients for performance comparisons on a wider population. The CHB-MIT dataset also allows for the state-of-the-art comparisons as it is a commonly used dataset to test seizure classifiers. Moreover, using these two datasets also measures how SOUL performs on datasets having different recording processes (iEEG versus scalp EEG).

The iEEG dataset was divided into 15% training, 15% validation, and 70% testing sets, as illustrated in Fig. 3. Contrary to random sampling during offline training, which is typically done in conventional machine learning approaches, time-series causality is maintained by considering only the first 30% (training + validation) of the data. Due to the limited seizure data for some patients in the CHB-MIT dataset, at least two seizure events were used for training and validation. However, if applicable, an approximate 15-15-70 split is still applied. For both datasets, the nonseizure samples were trimmed to balance the training data (equal number of seizure and nonseizure training points). Nonseizure samples closest to the start and end of the seizure events were retained to improve classification accuracy. The duration of the training, validation, and test sets, as well as the division between seizure and nonseizures samples, are shown in Table I.

TABLE I
DATASET PARTITIONING BETWEEN TRAINING, VALIDATION, AND TEST

| | Training set | | Validation set | | | Test set | | |
|---|---|---|---|---|---|---|---|---|
| | Seizures (#, sec) | | Seizures (#, sec) | | NS (hrs) | Seizures (#, sec) | | NS (hrs) |
| Patient 1* | 51 | 2479 | 51 | 2295 | 21.4 | 239 | 12667 | 98.2 |
| Patient 2* | 31 | 1953 | 31 | 2046 | 19.3 | 142 | 9514 | 190.1 |
| Patient 3* | 71 | 2982 | 71 | 3834 | 25.3 | 332 | 20252 | 246.9 |
| Patient 1 | 1 | 42 | 1 | 29 | 0.57 | 5 | 385 | 37.1 |
| Patient 2 | 1 | 84 | 1 | 83 | 1.06 | 1 | 11 | 19.2 |
| Patient 3 | 1 | 54 | 1 | 67 | 1.11 | 5 | 295 | 36.8 |
| Patient 4 | 1 | 51 | 1 | 113 | 8.37 | 2 | 222 | 129 |
| Patient 5 | 1 | 117 | 1 | 112 | 7.18 | 3 | 339 | 26.7 |
| Patient 6 | 2 | 33 | 2 | 39 | 10.0 | 6 | 101 | 54.6 |
| Patient 7 | 1 | 88 | 1 | 98 | 3.55 | 1 | 145 | 23.5 |
| Patient 8 | 1 | 173 | 1 | 192 | 3.06 | 3 | 564 | 16.2 |
| Patient 9 | 1 | 66 | 1 | 81 | 5.43 | 2 | 137 | 40.8 |
| Patient 10 | 1 | 37 | 1 | 72 | 16.2 | 5 | 352 | 16.1 |
| Patient 11 | 1 | 24 | 1 | 34 | 1.67 | 1 | 754 | 1.04 |
| Patient 12 | 6 | 175 | 6 | 258 | 4.68 | 28 | 1042 | 16.0 |
| Patient 13 | 2 | 118 | 2 | 99 | 7.89 | 8 | 342 | 7.83 |
| Patient 14 | 1 | 16 | 1 | 22 | 0.83 | 6 | 147 | 22.6 |
| Patient 15 | 3 | 286 | 3 | 244 | 4.78 | 14 | 1412 | 20.7 |
| Patient 16 | 2 | 22 | 2 | 24 | 5.03 | 6 | 58 | 8.69 |
| Patient 17 | 1 | 92 | 1 | 117 | 1.21 | 1 | 90 | 19.1 |
| Patient 18 | 1 | 52 | 1 | 32 | 0.18 | 4 | 245 | 6.47 |
| Patient 19 | 1 | 80 | 1 | 79 | 1.74 | 1 | 83 | 1.08 |
| Patient 20 | 1 | 31 | 1 | 32 | 1.37 | 6 | 247 | 17.2 |
| Patient 21 | 1 | 58 | 1 | 52 | 1.37 | 2 | 97 | 13.1 |
| Patient 22 | 1 | 60 | 1 | 76 | 4.94 | 1 | 74 | 9.11 |
| Patient 23 | 1 | 115 | 1 | 22 | 1.75 | 5 | 301 | 23.7 |
| Patient 24 | 2 | 54 | 2 | 65 | 2.62 | 11 | 397 | 18.0 |

* = iEEG dataset, S = Seizures, NS = Non-seizures

When testing the classifier, data are fed into the chip through a built-in scan chain, which also streams out the classifier output for checking. Classifier accuracy is measured by calculating the sensitivity (true positive rate) and specificity (true negative rate), as shown below

$$\text{Sensitivity} = \frac{\text{number of detected seizures (label1)}}{\text{total number of seizures}}$$
$$\text{Specificity} = \frac{\text{number of negative detections (label0)}}{\text{total number of non} - \text{seizures}}.$$

Both sensitivity and specificity are calculated on a sample-per-sample basis. That is, the corresponding classifier output for every sample is checked against the true label that was provided with the dataset. This method increases the granularity of the reported sensitivity and specificity, which is beneficial for datasets that contain very few seizures such as CHB-MIT.

TABLE II
COMPARISON TABLE VERSUS RECENT STATE-OF-THE-ART ON-CHIP SEIZURE EVENT CLASSIFIERS

|  | JETCAS 2018[21] | ISSCC 2018[13] | ISSCC 2020[22] | ISSCC 2020[14] | JSSC 2020[15] | ISSCC 2021[20] | VLSI 2021[16] | This Work | |
|---|---|---|---|---|---|---|---|---|---|
| Dataset (# of patients) | ieeg.org (20) | EPILEPSIAE (4) | EPILEPSIAE (-) (c) | CHB-MIT (23) | CHB-MIT + Local (24 + 2) | Bonn University (5) | CHB-MIT + Local (24 + 1) | iEEG (3) | CHB-MIT (24) |
| Channels | 32 | 32 | 8 | 8 | 16 | - (c) | 16 | 8 | |
| Classifier | Decision trees | EDM-SVM | EDM-Brain Forest | Coarse/Fine LS-SVM | Non-linear SVM | Reconfigurable Neural network | GTCA-SVM | Logistic regression + SGD (SOUL) | |
| Online Training Algorithm (Labeling method) | X | X | X | X | ADMM (Supervised) | ALE (Supervised) | GTCA (Supervised) | SGD (Unsupervised) | |
| Sensitivity (%) | 83.7 | 100 | 96.7 | 97.8 | 96.6 | 99.84 | 97.8 | 97.9 | 97.5 |
| Specificity (%) | 88.1 | - (a) | - (a) | 99.7 | 99.5 | - (c) | 99.5 | 98.2 | 98.2 |
| Latency (s) | 1.79 | <0.1 | - (c) | <0.3 | 0.71 | - (c) | <1 | 2.6 | 1.6 |
| Technology (nm) | 65 | 130 | 65 | 180 | 40 | 65 | 40 | 28 | |
| Energy Efficiency (nJ/cls) | 41.2 | 104,000 | 36 | 14,200 | 170,000 | 2,060 | 430 - 680 | 1.5 | |
| Classifier Area (mm²) | 1 | 5 (b) | 1 (b) | 3.51 | 4.5 | 1.74 | 2.25 (b) | 0.1 | |

a: Reported 0.80 false alarms per hour    b: Estimated from chip photo    c: Not reported

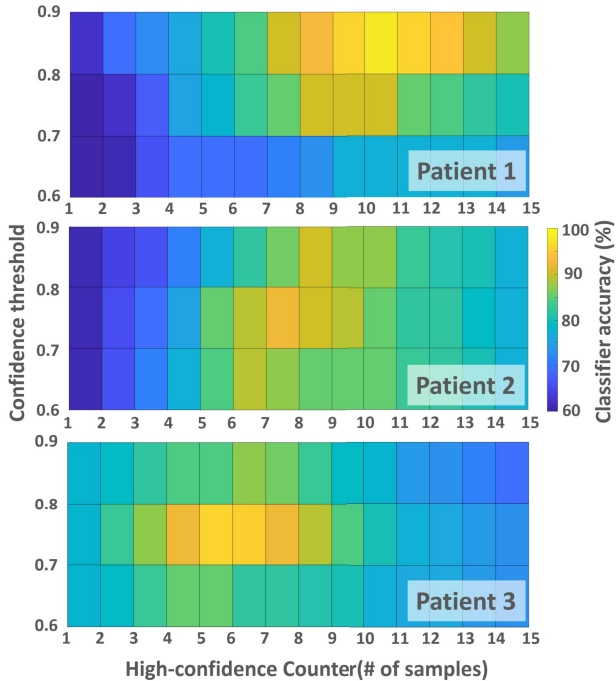

Fig. 6. Classifier accuracies (z-axis) during the tuning phase of the two hyperparameters: CT and HC counter. Different patients have different optimal values.

### A. Tuning the Online Learning Hyperparameters

As discussed in Section II, the HC and CT hyperparameters are tuned on a patient-specific basis. The values of these hyperparameters depend on both the short-term and long-term variability of the EEG signals per patient. Noisy EEG signals require higher CTs. Long-term time-varying signals require shorter HC so that the classifier can track signal changes faster. Fig. 6 shows the achieved accuracies of the classifier during the hyperparameter tuning on three patients from the iEEG dataset [8]. HC, measured in terms of the number of samples (each sample corresponds to one complete feature window which is 1 ms), was swept from 1 to 15. CT, which thresholds the logistic function output, was swept from 0.6 to 0.9.

Fig. 6 shows that for Patient 1, the optimal hyperparameter values are CT = 0.8 and HC = 10. The high CT value implies that the EEG signal is relatively noisy. Thus, the threshold needs to be high to avoid misclassifications negatively affecting the online training process. The optimal HC is also high to further mitigate the noise. Patient 3, on the other hand, has low hyperparameter values (CT = 0.7, HC = 5). These imply that the EEG signal is less noisy (lower CT) and that the signal varies over the long term (lower HC to stay on track). Fig. 6 also shows that without patient-specific tuning, a common value for the hyperparameters (CT = 0.7, HC = 7) can be used instead for these three patients, albeit with maximum sensitivities only reaching approximately 90%.

### B. Performance on Long-Term iEEG Data

The iEEG data are composed of recordings from three human patients that had the lowest seizure prediction performances out of the ten patients in a clinical trial done in [8]. Fig. 7 shows the classification performance over time of three classifiers: SOUL, logistic regression, and a representative SVM. The latter two are only trained offline. Incorporating online learning results in an average sensitivity and specificity of 97.9% and 98.2% for the three patients. For the three patients, the average sensitivity improvement is 6.5% with <1% specificity degradation. This degradation is a consequence of utilizing a linear classifier, such as the logistic regression used in SOUL. As a new seizure training point is introduced during retraining, the classifier tends to bias toward increased sensitivity (so that succeeding seizures can be better detected) while sacrificing specificity (as higher sensitivity leads to increased false alarms). This effect can be mitigated by also training on the nonseizure segments (further described in Section IV-B). The overall specificity degradation of <1% is considered an acceptable tradeoff for a more significant improvement in sensitivity.

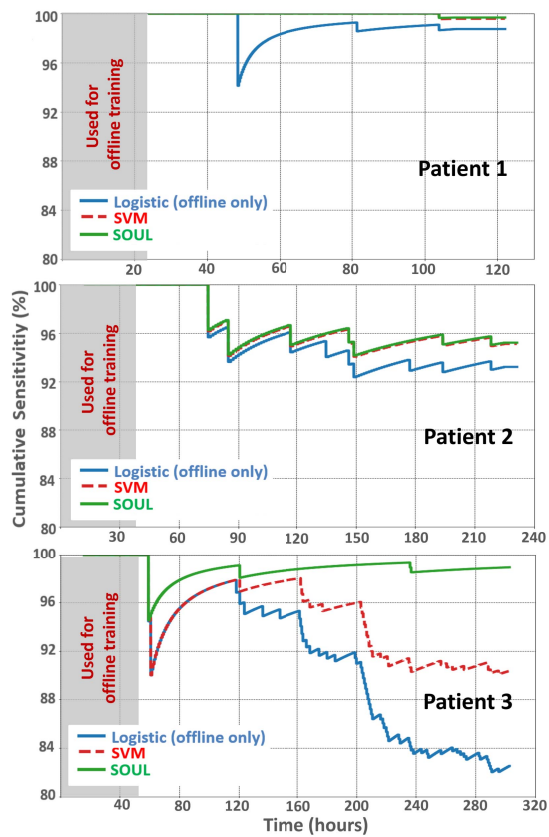

Fig. 7. Comparison of cumulative sensitivity over time for different classifiers versus SOUL for all three iEEG recordings. For patients 1 and 2, SOUL and SVM sensitivity performance is equivalent.

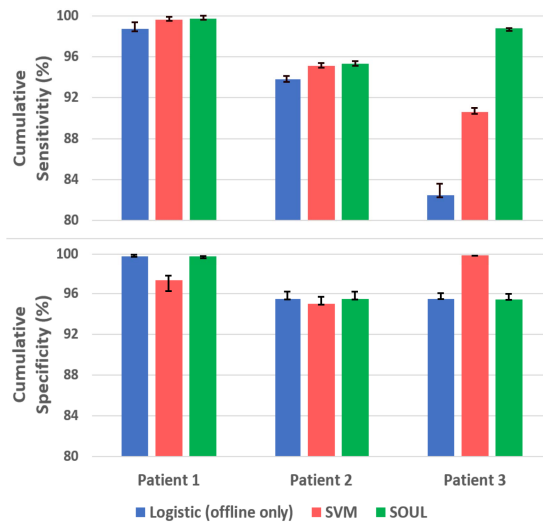

Fig. 8. Final sensitivity and specificity values at the end of the iEEG testing period; error bars indicate max and min values within the last 24 h. All classifiers are trained so that specificity remains >95%.

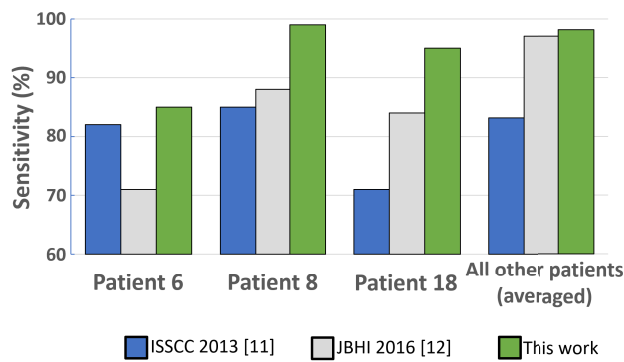

Fig. 9. Sensitivity comparison against state of the art on the CHB-MIT dataset on select patients where significant improvement was observed. For the remainder of the patients, an average of 1%–3% improvement was observed.

While the performance of SOUL and the SVM is the same for Patients 1 and 2, a significant performance difference is observed for Patient 3. The decreasing sensitivities of the conventional offline-only-trained classifiers demonstrate that seizure patterns change over time, which leads to missed detections. As SOUL tunes the feature weights during classification, it effectively tracks the iEEG signal variability, allowing sensitivity to be maintained over time. Fig. 8 shows the summary of final sensitivity and specificity values across all three patients after running the test dataset. In this work, <1.2 false alarms per day (>95% specificity) are maintained for all patients in all algorithms, equivalent to false alarm rates of commercial devices [4].

C. Performance on Scalp EEG Data

The CHB-MIT dataset consists of scalp EEG recordings from 24 pediatric subjects with intractable seizures [19]. Across all subjects, the mean recording time was 41 h and the mean number of recorded seizure events per subject was 7.6. Compared to the iEEG dataset, this is significantly shorter in terms of recording time and the number of seizures. However, this dataset is used for comparison since most seizure detection systems refer to this dataset.

Fig. 9 shows the comparison to other works which presented their results on a per-patient basis across all 24 subjects [11], [12]. For some select subjects (subjects 6, 8, 18), greater than 12% improvement in sensitivity was observed. For the rest of the subjects, there was a 1%–3% improvement. Across all 24 subjects, the average sensitivity improved by 14.8% compared to [11], and 1.8% compared to [12]. The average specificity for all subjects was 98.2%, which is 2.7% better than [11] and 0.2% better than [12]. Since SOUL was able to output the correct label within the seizure window for all seizure events, the event-based sensitivity for all patients is 100%.

The classification performance of SOUL on the two datasets shows that the proposed unsupervised online learning scheme works for both iEEG and scalp EEG. This demonstrates the flexibility of the algorithm on different EEG recording methods, as well as on different recording lengths. Compared to the other classifiers, SOUL maintains equal or higher sensitivities over the entire classification period. As classification goes on for longer, the sensitivity improvement from SOUL increases, as seen in Fig. 7 (Patient 3).

Fig. 10 shows the percentage of the test data translating to HC classifications. The HC percentage for seizure classification is very low (~5% for iEEG and <1% for CHB-MIT) with respect to the overall test data. This is directly correlated with the rarity of seizure events. Accordingly, the HC percentage for nonseizure classification is very high (~90%) as these comprise the bulk of the EEG recordings. The amount of retraining enabled from the consecutive HC classifications is also shown in the figure. It is more likely for SOUL to retrain on a seizure event (>50%) than on nonseizure data (<10%). This is due to how the HC hyperparameter for nonseizures is set up. As will be covered later, HC for nonseizures is 10× the HC value for seizures. This is to minimize the retraining frequency of SOUL over the long nonseizure periods. Effectively, SOUL biases toward higher sensitivity by retraining more frequently on the rare seizure events.

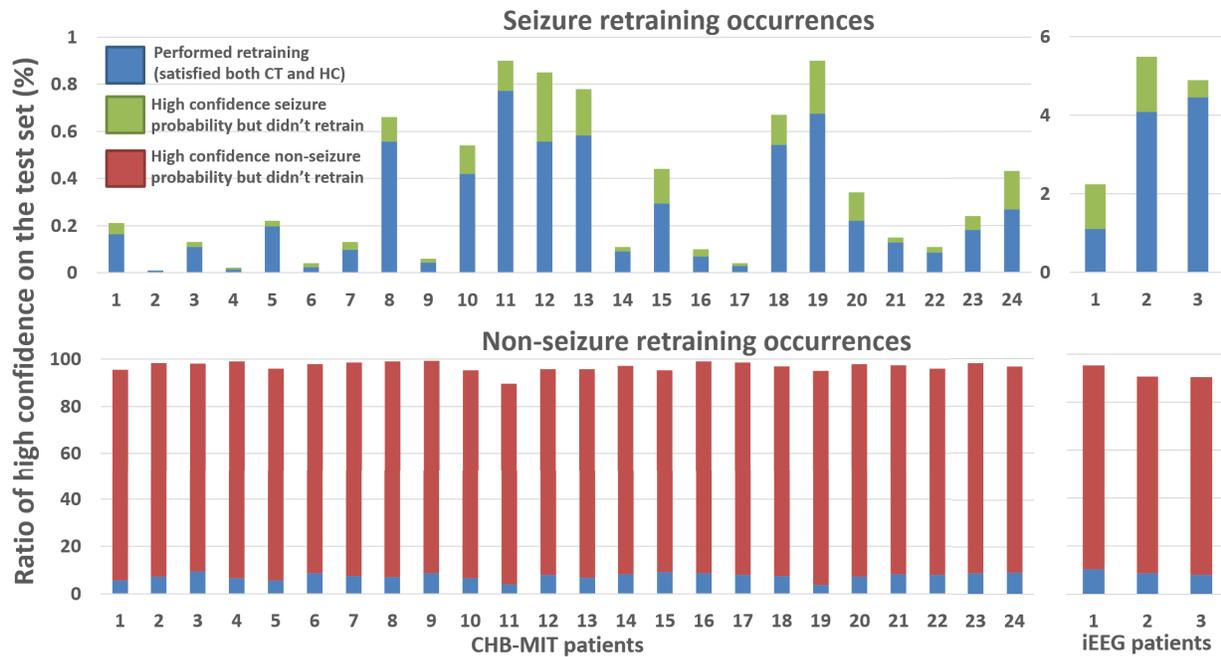

Fig. 10. Seizure and nonseizure retraining occurrences based on HC classifier outputs for the two datasets. SOUL retrains more during the rare seizure segments to bias toward increasing sensitivity (increasing seizure detection rate).

### D. Classifier Stability

SOUL was also tested for classification stability, which measures how a machine learning algorithm performs when the dataset is perturbed by noise. The accuracy of a stable classifier does not significantly change after perturbation since it should be able to generalize and not overfit on a given dataset. Artificial white noise was added to the two datasets used in this work. The standard deviation of the noise was swept from 1 to 10 $\mu V_{rms}$ at 1 $\mu V_{rms}$ increment. At 10 $\mu V_{rms}$, the added noise is comparable with the average biological noise measured during the nonseizure segments for each dataset. For each level, ten training and classification runs were performed to average out the effects of noise on the classification accuracy.

Fig. 11 shows the performance of SOUL when noise was added to the long-term iEEG data. This is then compared against the other classifiers shown in Fig. 7. The average sensitivity for the three patients was plotted. The sensitivity values vary within 1%–2% from the mean at every noise level, shown as error bars in Fig. 11. It can be observed that while all classifier accuracies degrade as the noise level increases, the average sensitivity of SOUL degrades much slower. At the maximum noise level, SOUL sensitivity decreased by only ~4% allowing it to achieve 11.6% better sensitivity than a representative SVM. This demonstrates both classification stability and the feasibility of the online learning scheme even with added noise on the iEEG dataset.

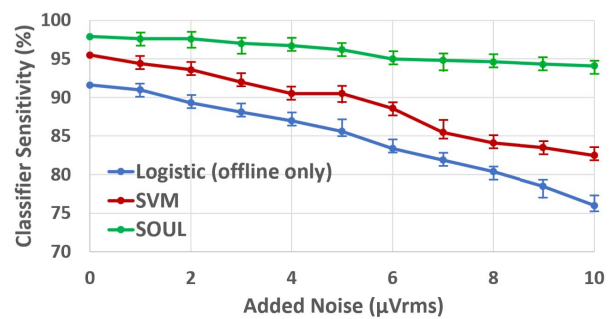

Fig. 11. Classifier sensitivities at increasing noise levels on the iEEG dataset, averaged per patient. Error bars represent the min and max values per run (ten runs per noise level).

The same setup was done on the CHB-MIT dataset. At 10 $\mu V_{rms}$ of noise, SOUL had an average sensitivity of 92.3% (down from 97.5% without noise) across all 24 patients. Each patient had >90% sensitivity and 100% event-based sensitivity. On a representative SVM, the average sensitivities decreased at a rate similar to the trend observed in Fig. 11. At the maximum noise level, the SVM sensitivity is down to 88.7%, which is 3.6% lower than SOUL. This demonstrates that SOUL was able to perform better than other representative classifiers when artificial noise is added in both iEEG and scalp EEG datasets. With the addition of 1–2 $\mu V_{rms}$, typical of the state-of-the-art neural recording frontends, the impact on SOUL sensitivity is <1%. This highlights the compatibility of SOUL as an online learning classification algorithm on an implantable device.

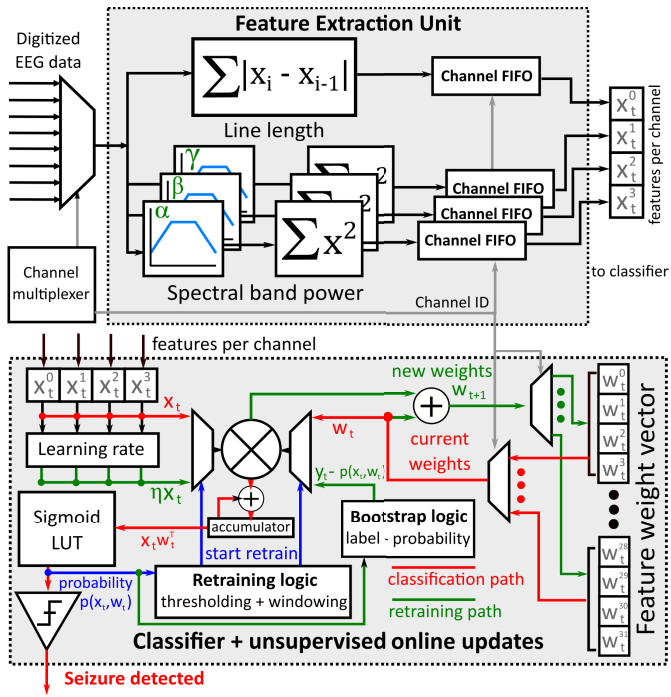

Fig. 12. System architecture (feature extraction unit + SOUL).

## IV. SYSTEM ARCHITECTURE

Fig. 12 shows the overall system architecture. The classifier receives 16-bit digitized neural data in eight channels clocked at 1 kS/s. The implemented system supports eight channels, but the algorithm is scalable to any number of channels.

Channel multiplexing and serialization are employed to minimize the duplication of hardware. The feature extraction unit is reused for each channel, which forces the system to run at 8× the sampling frequency (8 kHz). Feature values from all eight channels are passed to the classifier hardware, which will be discussed later in this section. As classifications are performed for every new sample (as also discussed in Section II-F), the classification rate of the system is 1 kHz.

### A. Feature Extraction Hardware

The feature extraction hardware unit is shown in Fig. 12 (upper half). Both line length and approximate spectral band power (using the sum of squares approximation) have a similar 100-sample register-based delay line, corresponding to the feature window, each connected to an accumulator to represent the summation of these 100 samples per feature. Each feature has a channel FIFO, controlled by the Channel ID signal that also controls the channel multiplexing state machine. Each channel FIFO is an eight-address register file that contains the current set of 100 samples for the corresponding active channel. The FIFO separates the feature data for each channel during the multiplexing phase.

The spectral band power block uses IIR filters instead of the conventional finite-impulse response (FIR) filter. During the feature extraction process, filters with at least 20 dB stopband were required for the spectral band power to work as a feature. If the filters do not meet those specifications, the approximated spectral band power using the sum of squares gets removed by the feature selection process leading to accuracy degradation of >10%. Designing FIR filters in MATLAB for a narrow passband, as an example, between 16 and 32 Hz, would require a minimum of 141 stages: each stage containing a register, an adder, and a multiplier. However, if elliptic IIR filters are used instead to achieve the same specification, it would only require three second-order sections: each section containing four registers, adders, and multipliers. Across all three spectral band power calculations, this filter choice translates to a 10× decrease in filter hardware requirements. When utilizing the elliptic filter architecture, the effects of frequency-dependent group delay on the classifier performance were ignored. It is assumed that this delay would be factored in during the offline training phase with minimal impact on detection latency.

The feature extraction unit computes in a 16-bit fixed-point format to avoid dedicated hardware for floating-point conversions. The Direct Form I IIR filter topology was used to avoid internal filter overflow. Given the 16-bit input to the system, 6 bits were set to be the integer part and the latter 10 were set to be the fractional part. This partitioning minimizes the round-off errors within the filter's internal states, which can cause instability. Through MATLAB filter design simulations, the number of bits can be reduced to 15 (with a 5–10 split between integer and fractional) given the datasets that were used for testing. However, 16 bits were retained as the hardware savings are marginal and an additional bit allows support for larger input signals to be processed.

### B. Classification and Online Learning Hardware

The SOUL hardware, shown in Fig. 12 (lower half), merges the two modes of operation of the classifier: classification mode and retraining mode. During the classification mode, shown as the red path in Fig. 12, the seizure probability is calculated using (1). The dot product for the logistic function is calculated on this mode. Since the four features from each channel are transferred one cycle at a time for every channel, the cumulative dot product is temporarily saved. Once all 4 × 8 features are collected, then classification will proceed.

The logistic function is approximated using a LUT to minimize computation hardware. While the classifier output is rounded-off to determine whether a seizure is detected or not, the accuracy for the LUT will matter since the value of the logistic function is part of the SGD feature weight update formula, as shown in Fig. 4. For this system, a ten-entry LUT was found to be enough, shown in Fig. 13, as it impacts the classifier accuracy by <1% compared to a classifier with full precision logistic function calculation.

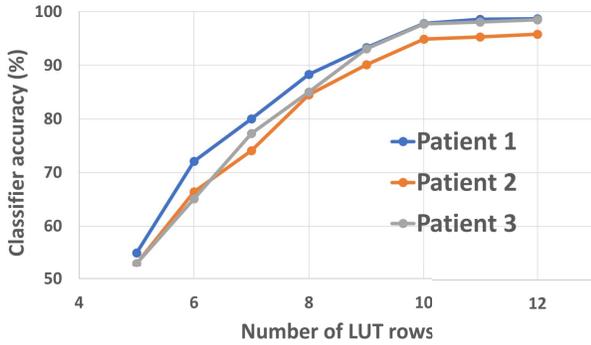

Fig. 13. Logistic LUT approximation impact on classifier accuracy.

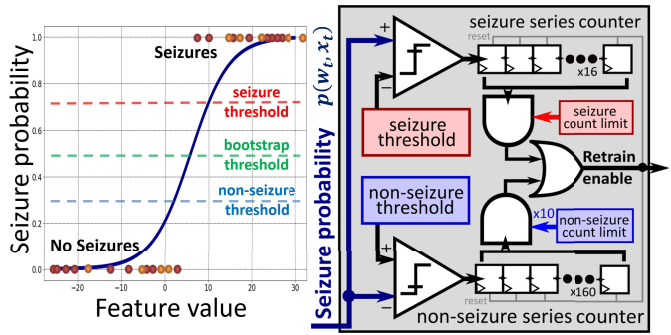

Fig. 14. Hardware implementation of the confidence thresholding and HC counters for robust online learning.

The output of the LUT provides the input to two sets of comparators, which correspond to the HC thresholds as described previously in Section II. The hardware for the confidence thresholding and the HC counters is shown in Fig. 14. Two separate CTs correspond to seizure and nonseizure. The seizure CT equals the value of CT, while the nonseizure CT equals $1 - \text{CT}$. The output of these comparators then goes to their corresponding series of shift registers representing the HC counters. The HC value for nonseizures is set to $10\times$ longer than the HC value for seizures to minimize the retraining frequency during the long nonseizure periods. This scaling balances the number of training points on the seizure and nonseizure events for an unbiased logistic model during the retraining period. Only when there is a series of HC probability outputs and either one of the HC counter limits is reached, the classifier goes into retraining mode. The retraining process can happen during either the seizure or nonseizure interval depending on which set of shift registers first reach the counter limit (corresponding to HC). The HC and CT parameters are programmable in hardware.

Fig. 12 (green path) shows the retraining mode calculations following the SGD formula for logistic regression. The learning rate for the retraining was set to be approximately 0.015 (1/64) and is calculated with simple right shifts. The bootstrap register computes the difference between the generated label (thresholded against 0.5 for unsupervised learning) and the actual LUT-based logistic function approximation. The retraining mode finishes in eight cycles, as the multiplier array is reused from the previous classification mode. During the update process, the old feature weight vectors are overwritten four at a time. Consequently, since the retraining mode consumes the same number of cycles as the classification, one input sample is ignored during the process. Once the retraining process is complete, the HC counters (shown at Fig. 14) reset, and the classification mode begins for the next input sample. Accordingly, the collection of HC detections starts again.

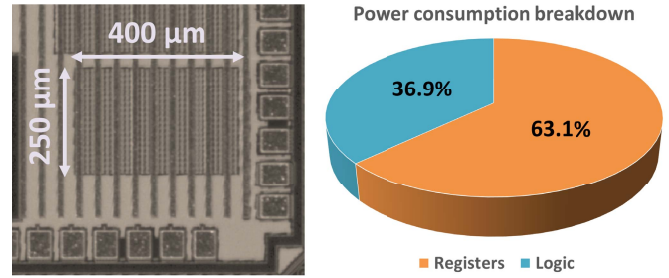

Fig. 15. Chip micrograph of SOUL in TSMC's 28 nm process and the power consumption breakdown (postlayout estimate).

## V. RESULTS AND DISCUSSION

The classifier was fabricated in Taiwan Semiconductor Manufacturing Company (TSMC's) 28 nm high-performance mobile (HPM) process occupying 0.1 mm$^2$ in area, shown in Fig. 15. The power consumption was measured to be 1.5 $\mu$W, which corresponds to an energy efficiency of 1.5 nJ/classification at 1 kHz classification rate. Since the power consumption is leakage-dominated in this regime, the described hardware reduction and reuse techniques significantly impact the total power. Further power reduction could be achieved by implementing the classifier in a low-power (LP) process, instead of the HPM variant that was used in this implementation. Digital logic and memory requirements were significantly reduced due to the relative computational simplicity of logistic regression coupled with architectural optimizations implemented to support online learning. The classifier-relevant memory is only 200 bytes, consisting of the feature weight values, the hyperparameter settings, and the HC counters. Moreover, as SOUL uses SGD and retrains in an unsupervised manner once deployed, only the current input features are used and stored at any given time. Consequently, there is no need to store a large amount of neural data for offline processing or training. When compared to the memory used for SVM-based systems [11]–[16], typically used to store signal data and classifier parameters, SOUL requires 300$\times$ smaller memory. A register-based memory implementation (instead of static random-access-memory (SRAM)-based) was used for SOUL, due to the very low-memory requirements. Nevertheless, the classifier-relevant memory is still very negligible when compared to the pipeline registers required for the filters and feature extraction logic, which dominate the power consumption as shown in Fig. 14. The architectural optimizations lead to at least 10$\times$ lower area and 24$\times$ better energy efficiency compared to other on-chip state-of-the-art classifiers. A comparison of this work with the state-of-the-art is shown in Table II.

The classifier performance in terms of accuracy has already been reported in Section III. It has been shown that SOUL

performed well on both short-term (CHB-MIT) and long-term (iEEG) datasets with differing recording qualities. Table II also shows the accuracy results together with the most recent on-chip state-of-the-art seizure detection systems. Some cited works used different datasets that introduced partiality since the EEG signal recording quality can be different.

For the CHB-MIT dataset, SOUL achieved slightly lower sensitivity and specificity values (0.3% and 1.5% lower, respectively) when compared to the more recent SVM-based classifiers [14]–[16]. This can be attributed to the limitation of a linear classifier, such as logistic regression used for SOUL, having difficulty with nonlinearly separable data. This is particularly highlighted for Patient 6, shown in Fig. 9, where the achieved sensitivity of SOUL is only 85%. Patient 6 has inconspicuous seizure events where the seizure and nonseizure samples look almost similar in the feature space. For this case, SOUL struggles to completely detect seizures (reducing average sensitivity) and raises false alarms during the nonseizure period (reducing average specificity). SVMs, on the other hand, do not suffer from this problem as they can be implemented as a nonlinear classifier (as highlighted back in Section II-B).

It is worth noting, however, that for recordings that do not have a very good linear separability between seizures and nonseizures, the unsupervised online learning scheme is relatively robust. In this case, samples that are hard to classify would typically have logistic function output probabilities very near to 0.5, which translates to low confidence. As shown in Section III-A, CT values are generally in the 0.7–0.8 range to represent HC. If the output probabilities are close to 0.5, then the retraining process cannot begin as a series of HC is not observed. Consequently, a nonlinear classification problem will not retrain SOUL in the wrong direction.

The classifier performance on long-term data (recording times ranging from several days to weeks) was not explicitly addressed in the other works. Many algorithms that work well on short EEG recordings (i.e., within a day) may fail to work on longer recordings. As SOUL has demonstrated that maintaining high accuracies over long periods of time is possible through SGD-based online learning, it would be interesting to see how the different online learning techniques from other implementations [15], [16], [20] would compare on the same long-term data.

The main advantage of SOUL over the other online learning classifiers in the literature is its unsupervised online learning feature. Providing external labels to constantly tune the classifier model parameters can be costly and impractical, especially when the system is already deployed. On the other hand, an unsupervised online learning scheme can enable a seizure detection system that would not require any external intervention to maintain high accuracies over time.

The reported seizure detection latency for SOUL was 1.6–2.6 s which is relatively high when compared to the state-of-the-art. This can be attributed to the frequency-dependent group delay introduced by IIR filters on the feature extraction unit, which varies the spectral power feature values when it arrives at the classifier. This group delay can be compensated by cascading a corresponding phase equalizer after every IIR filter, which increases the filter hardware requirements by approximately $2\times$. The relatively high detection latency might also be a consequence of the limited feature set that was used since the feature selection process only selected features based on accuracy and not latency. Nevertheless, it has been shown [3] that latencies less than 5 s have demonstrated clinical efficacy in detection-triggered stimulation devices.

This work has demonstrated an unsupervised, online learning classifier that can outperform conventional classifiers in terms of seizure detection accuracy, especially in the long term, while still being significantly more energy efficient. The achieved low-energy consumption was a direct consequence of both the algorithm choice and the architectural optimizations. Selecting logistic regression as the base classifier for SOUL enabled significantly low-memory overhead compared to the conventional SVM-based classifier. While logistic regression, on its own, may not perform well on datasets that are not linearly separable, incorporating SGD (also a relatively simple calculation for logistic regression) enabled dynamic feature weight updates. This technique allowed SOUL to maintain high accuracies even as seizure patterns change, avoiding the need for a relatively complex, high-memory overhead classifier to capture such changes. Architectural optimizations also reduced the hardware requirements, leading to a significantly less overall area and leakage power. Indeed, reducing the digital backend power consumption might not offer a significant benefit in terms of overall system power when the analog front ends are included. However, given that the current implementation is significantly more energy-efficient than the state-of-the-art, this gives more room for more complex feature extraction units to be incorporated with the classification hardware. This can further improve the classifier performance especially on nonlinearly separable data, as well as long-term performance of the unsupervised online learning scheme.

## VI. CONCLUSION

To the authors' knowledge, this work is the first to demonstrate an on-chip, unsupervised, online learning classifier for seizure detection. The classifier can continuously update, without external intervention, from an initial offline-trained model using a combination of SGD and bootstrapping. This allows high seizure detection accuracies to be maintained over longer periods of time when compared to static offline-only-trained classifiers. The classifier performance has been evaluated on two datasets, for a total of 27 human subjects. For the long-term iEEG dataset, incorporating online learning results in an average sensitivity and specificity of 97.9% and 98.2%, respectively, improving sensitivity by 6.5% on average with <1% specificity degradation over three patients. For the scalp EEG dataset, the classifier achieves 97.5% and 98.2% average sensitivity and specificity over 24 subjects. The sensitivity for the subjects either stayed the same (6/24) or improved (15/24) by 1%–3%. Moreover, an improvement of >12% was observed on three subjects when compared against other state-of-the-art presenting a per-subject sensitivity breakdown.

A significant benefit of this online learning approach is that the reported high accuracies were achieved on energy-efficient hardware. The combination of the proposed algorithmic approach and circuit-level optimizations resulted in an energy efficiency of 1.5 nJ/classification, which is at least 24× better than the state-of-the-art. The unsupervised classifier continuously improves its model over time and can dynamically adapt to neural pattern changes in real time, mitigating the need for in implantable or wearable seizure detection systems.

ACKNOWLEDGMENT

The authors would like to thank to the sponsors of the Berkeley Wireless Research Center and also like to thank the Savio computational cluster resource provided by the Berkeley Research Computing Program. They would also like to thank Dr. Mark Cook and Dr. Dean Freestone for providing the intracranial electroencephalography (iEEG) patient dataset.

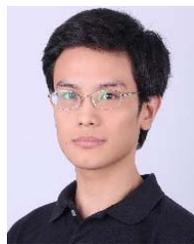

**Adelson Chua** (Graduate Student Member, IEEE) received the B.S. degree in computer engineering and the M.S. degree in electrical engineering from the University of the Philippines Diliman, Quezon City, Philippines, in 2012 and 2014, respectively. He is currently pursuing the Ph.D. degree with the Department of Electrical Engineering and Computer Sciences, University of California at Berkeley, Berkeley, CA, USA, with a focus on integrated circuits, computer architecture, and machine learning.

He was a key member of several Government-Funded Projects through the Department of Science and Technology, which includes the implementation of an error-detecting and correcting microprocessor; creation of a vision-capable microcontroller; and design of a wide temperature range, powerline, energy harvesting sensor node system on a chip. He has also been a part of the Philippine-California Advanced Research Institutes (PCARI) Project of the Philippine Government's Commission on Higher Education in collaboration with California-based and leading Philippine academic institutions. He is currently a member of the Faculty of Microelectronics, Electrical and Electronics Engineering Institute, University of the Philippines Diliman. He is a Graduate Student affiliate with the Berkeley Wireless Research Center (BWRC), Berkeley. His research is on on-chip machine learning and energy-efficient computing for implantable neural interfaces.

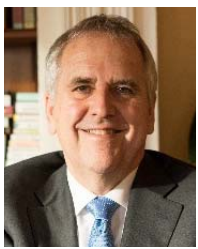

**Michael I. Jordan** (Fellow, IEEE) received the master's degree in mathematics from Arizona State University, Tempe, AZ, USA, in 1980, and the Ph.D. degree in cognitive science from the University of California San Diego, San Diego, CA, USA, in 1985.

He was a Professor with the Massachusetts Institute of Technology (MIT), Cambridge, MA, USA, from 1988 to 1998. He is currently a Pehong Chen Distinguished Professor with the Department of Electrical Engineering and Computer Science and the Department of Statistics, University of California at Berkeley, Berkeley, CA, USA. His research interests bridge the computational, statistical, cognitive, biological, and social sciences.

Prof. Jordan is a member of the National Academy of Sciences, the National Academy of Engineering, the American Academy of Arts and Sciences, and a foreign member of the Royal Society. He is a fellow of the American Association for the Advancement of Science. He is a fellow of the AAAI, ACM, ASA, CSS, IMS, ISBA, and SIAM. He was a Plenary Lecturer at the International Congress of Mathematicians in 2018. He was a recipient of the IMS Medallion Lecture in 2004, ACM/AAAI Allen Newell Award in 2009, the IMS Neyman Lecture in 2011, the David E. Rumelhart Prize in 2015, the IJCAI Research Excellence Award in 2016, the IEEE John von Neumann Medal in 2020, the Ulf Grenander Prize from the American Mathematical Society in 2021, and the Inaugural IMS Grace Wahba Lecture in 2022.

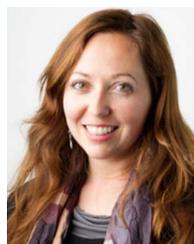

**Rikky Muller** (Senior Member, IEEE) received the B.S., and M.Eng. degrees from the Massachusetts Institute of Technology (MIT), Cambridge, MA, USA, in 2004, and the Ph.D. degree from the University of California at Berkeley (UC Berkeley), Berkeley, CA, USA, in 2013, all in electrical engineering and computer science.

She was a McKenzie Fellow and a Lecturer of electrical engineering with The University of Melbourne, Melbourne, VIC, Australia. She is currently an Assistant Professor of electrical engineering and computer sciences with UC Berkeley, where she holds the S. Shankar Sastry Professorship in emerging technologies. She is currently the Co-Director with the Berkeley Wireless Research Center (BWRC), Berkeley, a Core Member with the Center for Neural Engineering and Prostheses (CNEP), Berkeley, and an Investigator with Chan-Zuckerberg Biohub, San Francisco, CA, USA. She was previously an IC Designer with Analog Devices, Cambridge, MA, USA, and was the Co-Founder of Cortera Neurotechnologies, Inc., Berkeley (acq. 2019) a medical device company focused on closed-loop deep brain stimulation technology. She was named one of MIT Tech Review's 35 Global Innovators under 35 (TR35) and Boston MedTech's 40 Healthcare Innovators under 40. Her research group focuses on emerging implantable and wearable medical devices and in developing low-power, wireless microelectronic and integrated systems for neurological applications.

Dr. Muller is a Bakar Fellow, a Hellman Fellow, and a member of the SSCS Advisory Committee, the Solid State Circuits Directions Committee, and Women in Circuits. She was a recipient of the National Academy of Engineering Gilbreth Lectureship, the NSF CAREER Award, the Keysight Early Career Professorship, the McKnight Technological Innovations in Neuroscience Award, and the IEEE Solid-State Circuits Society New Frontier Award. She is the IMMD Subcommittee Chair for IEEE ISSCC and has previously served on the Technical Program Committees of IEEE CICC and BioCAS. She is a Distinguished Lecturer for the Solid-State Circuits Society. She served as a Guest Editor for the IEEE JOURNAL OF SOLID-STATE CIRCUITS.